\documentclass[luseAMS,usenatbib]{mnras}

\usepackage{newtxtext,newtxmath}

\usepackage[T1]{fontenc}
\usepackage{ae,aecompl}

\usepackage{graphicx}	
\usepackage{amsmath}	
\usepackage{amssymb}	

\usepackage{xcolor}

\usepackage{natbib}

\usepackage[colorinlistoftodos]{todonotes}
\usepackage[normalem]{ulem}

\title[Search EM counterpart by TOROS to GW events in O2]{TOROS Optical follow-up of the Advanced LIGO-VIRGO O2 second observational campaign}

\author[R. Artola et al.]{Rodolfo Artola,$^{1}$ %
Martin Beroiz,$^{2}$ %
Juan Cabral,$^{1}$ %
Richard Camuccio,$^{2}$ %
Moises Castillo,$^{2}$ \newauthor %
Vahram Chavushyan,$^{3}$ 
Carlos Colazo,$^{1}$ %
Hector Cuevas Larenas,$^{4}$ %
Darren L. DePoy,$^{5}$ \newauthor%
Mario C. D\'{\i}az,$^{2}$  %
Mariano Dom\'{\i}nguez,$^{1}$  %
Deborah Dultzin,$^{6}$  %
Daniela Fern\'{a}ndez,$^{7}$  \newauthor %
Antonio C. Ferreyra,$^{1}$  %
Aldo Fonrouge,$^{2}$ %
Jos\'{e} Franco,$^{6}$ %
Dar\'{\i}o Gra\~{n}a,$^{1}$ \newauthor %
Carla Girardini,$^{1}$ %
Sebasti\'an Gurovich,$^{1}$   %
Antonio Kanaan,$^{8}$ %
Diego G. Lambas,$^{1}$ \newauthor %
Marcelo Lares,$^{1}$  %
Alejandro F. Hinojosa,$^{2}$  %
Andrea Hinojosa,$^{2}$ %
Americo F. Hinojosa,$^{2}$ \newauthor %
Omar L\'{o}pez-Cruz,$^{3}$ 
Lucas M. Macri,$^{5}$  
Jennifer L. Marshall,$^{5}$ 
Raul Melia,$^{1}$ \newauthor %
Wendy Mendoza,$^{2}$ %
Jos\'{e} L. Nilo Castell\'on$^{4, 9}$, %
Nelson Padilla,$^{7}$ 
Victor Perez,$^{2}$  \newauthor %
Tania Pe\~{n}uela,$^{2}$ %
Wahltyn Rattray,$^{1}$ %
V\'{\i}ctor Renzi,$^{1}$  %
Emmanuel R\'{\i}os-L\'{o}pez,$^{3}$ \newauthor 
Amelia Ram\'{\i}rez Rivera,$^{4}$ %
Tiago Ribeiro,$^{10}$ %
Horacio Rodriguez,$^{1}$ %
Bruno S\'{a}nchez,$^{1}$  \newauthor %
Mat\'{\i}as Schneiter,$^{1}$ %
William Schoenell,$^{11}$ %
Manuel Starck,$^{1}$ %
Rub\'en Vrech,$^{1}$ \newauthor%
Cecilia Qui\~{n}ones,$^{1}$  %
Luis Tapia,$^{1}$ %
Marina Tornatore,$^{1}$ %
Sergio Torres-Flores,$^{4}$ \newauthor %
Ervin Vilchis,$^{2}$ %
Adam Zadro\.{z}ny$^{2,12}$\thanks{E-mail: adam.zadrozny@utrgv.edu (AZ)} \\
$^{1}$Universidad Nacional de C\'{o}rdoba, IATE, C\'ordoba, Argentina \\
$^{2}$Center for Gravitational Wave Astronomy, The University of Texas Rio Grande Valley, Brownsville, TX, USA\\
$^{3}$Instituto Nacional de Astrof\'{\i}sica, \'{O}ptica y Electr\'{o}nica, Tonantzintla, Puebla, M\'{e}xico \\
$^{4}$Departamento de F\'{\i}sica y Astronom\'{\i}a, Universidad de La Serena, La Serena, Chile\\
$^{5}$George P.~and Cynthia W.~Mitchell Institute for Fundamental Physics \& Astronomy, Department of Physics \& Astronomy,\\ Texas~A\&M~University, College Station, TX, USA\\
$^{6}$Instituto de Astronom\'{\i}a, Universidad Nacional Aut\'onoma de M\'{e}xico, AP 70-264, 04510 M\'{e}xico D.F., Mexico \\
$^{7}$Instituto de Astrof\'{\i}sica, Pontificia Universidad Cat\'{o}lica de Chile, Santiago, Chile \\
$^{8}$Universidade Federal de Santa Catarina, Florian\'{o}polis, Brazil \\
$^{9}$Instituto de Investigaci\'on Multidisciplinario en Ciencia y Tecnolog\'{\i}a, Universidad de La Serena, La Serena, Chile\\
$^{10}$Large Synoptic Survey Telescope, Tuscon, AZ, USA \\
$^{11}$Universidade Federal do Rio Grande do Sul, Porto Alegre, Brazil\\
$^{12}$National Centre for Nuclear Research, Astrophysics Division, ul. Ho\.{z}a 69, 00-681 Warsaw, Poland
}
\date{Accepted XXX. Received YYY; in original form ZZZ}

\pubyear{2018}

\begin{document}
\label{firstpage}
\pagerange{\pageref{firstpage}--\pageref{lastpage}}
\maketitle

\begin{abstract}
We present the results of the optical follow-up, conducted by the TOROS collaboration, of gravitational wave events detected during the Advanced LIGO-Virgo second observing run (Nov 2016 -- Aug 2017). Given the limited field of view ($\sim100\arcmin$) of our observational instrumentation we targeted galaxies within the area of high localization probability that were observable from our  sites. We analyzed the observations using difference imaging, followed by a Random Forest algorithm to discriminate between real and bogus transients. For all three events that we respond to, except GW170817, we did not find any bona fide optical transient that was plausibly linked with the observed gravitational wave event. Our observations were conducted using telescopes at Estaci\'{o}n Astrof\'{\i}sica de Bosque Alegre, Cerro Tololo Inter-American Observatory, and the Dr. Cristina V. Torres Memorial Astronomical Observatory. Our results are consistent with the LIGO-Virgo detections of a binary black hole merger (GW170104) for which no electromagnetic counterparts were expected, as well as a binary neutron star merger (GW170817) for which an optical transient was found as expected.
\end{abstract}




\section{Introduction}

The network of advanced ground-based gravitational wave (GW) interferometers consisting of the LIGO observatories \citep{LSC2015} and the VIRGO observatory \citep{Acernese2015} conducted their second observing run (O2) between November 2016 and August 2017. The detectors are designed to be capable of detecting GWs emitted by the mergers of compact objects like binary neutron star (BNS) systems, binary black hole (BBH) systems, or black hole-neutron star (BHNS) systems, out to distances of hundreds of Mpc \citep[see][and references therein]{Abbott2016b}. 

It is expected that if the mergers of compact objects contain at least one neutron star, electromagnetic (EM) radiation will be emitted during the event. Different EM counterparts, arising due to origins in expanding r-process ejecta and interaction with the surrounding stellar environment, could range from very short duration gamma-ray bursts (GRBs) and X-rays to longer duration emission at optical, near infrared, and radio wavelengths \citep[][]{Li1998,Nakar2011,Metzger2012,Barnes2013,Berger2014,Cowperthwaite2015}. Simultaneous detection of a merger event by GW and EM observatories provides an integrated astrophysical interpretation of the event and is instrumental in producing better estimates for the distance and energy scales of the event as well as improving the estimation of its orbital parameters.

The TOROS - ``Transient Optical Robotic Observatory of the South'' -  \citep[TOROS;][]{Benacquista2014} Collaboration  was  formed motivated by the desire to participate in these observations. TOROS  seeks to ultimately deploy a wide-field optical telescope on Cord\'on Mac\'on in the Atacama Plateau of northwestern Argentina \citep{Renzi2009,Tremblin2012}. Independently of the pursuit of this goal the collaboration had access to other astronomical resources which will describe immediately below.
Motivated by these goals, TOROS signed on April 5, 2014, a memorandum of understanding with the LIGO VIRGO collaboration (LVC) and participated in its first observing run (O1) from Sep 2015 -- Jan 2016 and the second observing run (O2) from Nov 2016 -- Aug 2017. Over the period of O2, TOROS used three facilities: the 1.5-m telescope at the Estaci\'on Astrof\'{\i}sica Bosque Alegre (EABA), the 0.826-m T80-South (T80S) at the Cerro Tololo Inter-American Observatory, and the 0.4-m telescope at the Dr. Cristina V. Torres Memorial Astronomical Observatory (CTMO).  Additionally, during the 2017A semester, TOROS had access to the Gran Telescopio Canarias (GTC) a segmented 10m telescope located at La Palma in the Canary Islands, we were assigned 5 hours as target of opportunity by the Mexican Time Allocation Committee (TAC).  The Mexican chapter of TOROS had access to the 2.1m telescope at the Guillermo Haro Observatory (OAGH) in Cananea Sonora. 

The first GW event, GW150914, a merger of a binary black-hole (BBH) system, was detected at the beginning of the advanced detectors first science run \citep{Abbott2016a}. The TOROS Collaboration was one of 25 teams that participated in the search for an EM counterpart in the southern hemisphere \citep{Diaz2016}. None of the participating teams found any optical transient associated with the event \citep{Abbott2016d}, a result consistent with theoretical predictions that BBH mergers should not produce optical counterparts.  

During the second observing run, O2, on  August 17, 2017 at 12:41:04.4 UT, the Advanced LIGO and Virgo detectors observed a high-significance GW event candidate designated GW170817 \citep{Abbott2017}, 
which was determined to be the first detection of a BNS merger. An EM counterpart was subsequently observed and designated AT 2017gfo and GRB170817A.  The TOROS collaboration  contributed to the follow up of this event as well \citep{Diaz2017}. 

All of the gravitational wave events detected in O1 and O2 science runs are described in GWTC-1: A Gravitational-Wave Transient Catalog of Compact Binary Mergers  \citep{LVC2018} published by the LIGO and Virgo Scientific Collaboration. Up till 30 Nov 2018 there are 11 gravitational wave events known. Ten of them are binary black-hole mergers and one, GW170817, is a binary neutron star merger. 

This paper is organized as follows: \S\ref{observations} discusses target selection and observations; \S\ref{data-analysis} describes the data reduction, difference imaging algorithms, and the classification of bogus and real transients; \S\ref{results} presents our results; \S\ref{conclusions} summarizes our findings.

\section{Observations}\label{observations}

\subsection{Alerts}

During the O2 Science Run, the TOROS network responded to three alerts: GCN circulars GCN\#268556, later designated GW170104; GCN\#G270580 that was later not classified as a GW; and GCN\#298048, later promoted to GW170817. GW170104 was followed up by EABA and it was imaged 1 to 2 weeks (between 9 and 17 days) after the event. GCN\#G270580 was followed by EABA, but the quality of the images obtained  was too poor to permit a useful  analysis.  GW170817 was followed up by the EABA and T80-South facilities. An optical counterpart was observed and lightcurves were obtained \citep{Diaz2017, Abbott2017}. 

\subsection{Telescope Pointing Strategy}

To search for EM counterparts we used a galaxy based approach. For each potential gravitational wave event detected, LIGO-Virgo issued a position probability map for the origin of the event. 
Even for 90\% confidence regions, the probability map usually covered a large area, typically much larger than $100\ sq.\ deg$. 
It is not practical to scan the whole higher probability localization area using a small FoV telescopes, like with FoV of $~0.5$ sq. deg that were available to TOROS during O2. 
Instead we assumed a compromise between additional pre-selection biases and viable increase in recovery rate, and also between the additional time for new pointing and the short  timescale of transient events associated with GWs. Consequently we decided to target only massive galaxies in the highest probability sky regions.
This  is not a limitation  though, since  the  probability of occurrence of NS or BH mergers is higher in close proximity to galaxies with considerable mass.
In fact, \citep{Hanna2014} 
showed that the use of galaxy catalogs can improve success rates by 10\% and up to 3 times compared to not utilizing these kind of catalogs. The discovery of the first joint GW-EM emission in close proximity to NGC 4993 confirmed the validity of this approach.

In implementing this strategy we compared the probability map for the origin of the gravitational wave event against objects in the Gravitational Wave Galaxy Catalog (GWGC, \citet{White2011}) and we chose the top 20 galaxies ranking them by their probability to be connected to the event. We also used additional cuts for target galaxies parameters and their position above the horizon. Impact of those cuts on search results is discussed in subsection \ref{sec:strategy}. 
One of our filters for observability was too conservative setting the limits on observable RA, making NGC 4993 fall outside the target list and was not initially selected by our pipeline as an observable target. For that reason,  NGC 4993 was not imaged on the first night. Nonetheless, it was observed the following night, after the optical transient (OT) coordinates were shared with LV-EM collaboration.

\begin{table*}
\caption{Targeted host galaxies for GW170104\label{table1}.\ \newline
Notes: ($^1$) local date of observation; ($^2$) from \citet{White2011}}
\begin{tabular}{llrrrr}
Date$^{1}$ & ID$^{2}$ & RA & Dec. & Total & D \\
&  & [hh:mm:ss] & [d:mm:ss] & $t_{\rm exp}$ & [Mpc] \\
2017-01-13 & PGC073926  & 01:20:36.56 & -36:05:34.40 & 41m & 79.03 \\
2017-01-13 & NGC1341    & 03:27:58.39 & -37:09:00.22 & 51m & 19.86 \\
2017-01-13 & NGC1808    & 05:07:42.31 & -37:30:46.66 & 1h 7m & 12.30 \\
2017-01-13 & ESO487-003 & 05:21:50.69 & -23:57:04.79 & 6m & 51.57 \\
2017-01-13 & ESO364-014 & 05:55:01.31 & -36:56:49.38 & 49m 30s & 79.22 \\
2017-01-14 & NGC1341    & 03:27:58.39 & -37:09:00.22 & 5m & 19.86 \\
2017-01-14 & ESO202-009 & 04:21:03.64 & -48:18:14.87 & 47m 25s & 61.29 \\
2017-01-14 & PGC147285  & 05:16:47.93 & -36:04:07.79 & 22m & 58.22 \\
2017-01-15 & ESO242-018 & 00:37:06.17 & -46:38:38.83 & 50m & 48.31 \\
2017-01-15 & NGC1567    & 04:21:08.75 & -48:15:17.39 & 50m & 60.72 \\
2017-01-15 & ESO425-010 & 06:08:57.37 & -27:48:07.45 & 44m & 36.32 \\
2017-01-15 & PGC3080859 & 05:19:35.94 & -36:49:02.50 & 46m & 61.86 \\
2017-01-15 & IC2143     & 05:46:52.64 & -18:43:34.82 & 45m & 38.02 \\
2017-01-15 & ESO555-005 & 05:51:39.85 & -18:01:21.90 & 44m 30s & 39.99 \\
2017-01-15 & ESO555-022 & 06:01:08.00 & -21:44:19.28 & 42m & 20.80 \\
2017-01-20 & PGC1312883 & 09:20:01.79 & +07:03:22.10 & 1h & 48.64 \\
2017-01-20 & UGC04959   & 09:20:32.14 & +07:04:26.90 & 1h & 75.08 \\
2017-01-21 & NGC1341    & 03:27:58.39 & -37:09:00.22 & 1h 14m & 19.86 \\
2017-01-21 & PGC025197  & 08:58:12.04 & -06:11:57.12 & 1h 1m 30s & 65.11\\
\end{tabular}
\end{table*}

\begin{table*}
\caption{Targeted host galaxies for GW170817\label{table2}.\ \newline
Notes: ($^1$) local date of observation; ($^2$) from \citet{White2011}}
\begin{tabular}{clrrrrr}
Date$^{1}$ & ID$^{2}$ & RA & Dec. & Total & D & Tile \\
 &  & [hh:mm:ss] & [d:mm:ss] & $t_{\rm exp}$ & [Mpc] & \# \\
2017-08-17 & PGC141857 & 14:10:33.49 & -52:19:01.42 & 15m & 41.722 & 1 \\
2017-08-17 & ESO221-030 & 14:10:41.12 & -52:11:02.90 & 15m & 42.319 & 1 \\
2017-08-17 & PGC448694 & 14:10:37.02 & -52:06:05.90 & 15m & 56.194 & 1 \\
2017-08-17 & PGC141859 & 14:20:23.53 & -55:04:06.64 & 8m & 36.667 & 2 \\
2017-08-17 & PGC2800412 & 14:17:10.00 & -55:37:11.64 & 8m & 52.583 & 2 \\
2017-08-17 & PGC166323 & 14:04:34.10 & -52:41:49.99 & 3m & 49.417 & 3 \\
2017-08-17 & PGC166327 & 14:05:56.62 & -53:43:10.42 & 3m & 52.500 & 3 \\
2017-08-17 & PGC463082 & 14:03:29.27 & -50:46:37.60 & 4m & 45.792 & 4 \\
\end{tabular}
\end{table*}

\begin{figure}
\includegraphics[width=0.5\textwidth]{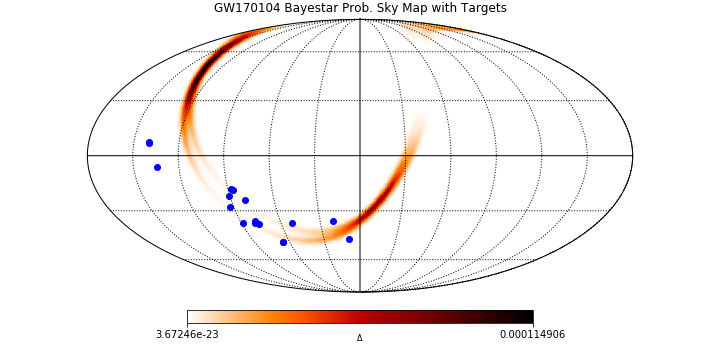}
\caption{Bayestar probability all-sky map for GW170104 with our targets in blue dots.\label{fig:gwtargets}}
\end{figure}

\ \par

\section{Data analysis}\label{data-analysis}

The initial data reduction followed the standard steps of bias and dark subtraction, flat-fielding using twilight sky frames, and illumination correction. We used Astrometry.net to perform the astrometry \citep{Lang2010}.We apply a difference image analysis  to search for transients, as is described in subsection \ref{2kk}.  To identify a potential optical transient we have used SExtractor \citep{Bertin1996}. There were two photometric pipelines used. One based on DAOPHOT \citep{Stetson1987} for analyzing the transients found and a second one based on SExtractor.

\ \par

\subsection{Photometry for GW170817}
\footnote{This section is a summary of the photometry section from the TOROS original paper describing the observation of GW170817. For more details please refer to \citep{Diaz2017}}
In the  analysis of GW170817 we used photometry obtained by the T80S and EABA 1.5m telescopes. We also utilized the IRAF suite \footnote{IRAF is distributed by the National Optical Astronomy Observatory, which is operated by the Association of Universities for Research in Astronomy (AURA) under cooperative agreement with the National Science Foundation.} {\tt CCDPROC} package to de-bias and flat-field the raw frames. We carried out time-series point-spread function (PSF) photometry using {\tt DAOPHOT/ALLSTAR} \citep{Stetson1987}, ALLFRAME \citep{Stetson1994} and related programs, kindly provided by P.~Stetson. The steps we performed closely follow those outlined in \citet{Macri2006} and \citet{Macri2015}. We first used {\tt DAOPHOT} to detect sources in each image with a significance of $4\sigma$ or greater, to identify bright isolated stars, and then determine their PSFs. We then used {\tt ALLSTAR} to obtain preliminary PSF photometry for all of the sources. Next, we used {\tt DAOMATCH} and {\tt DAOMASTER} to derive robust geometric transformations between all of the images obtained with a given telescope and filter, and create master frames. Given the close proximity of the transient to its host galaxy, we used the {\tt IMFIT} package \citep{Erwin2015} to model and subtract its light distribution from each image twice. 
We performed the aforementioned procedures with {\tt DAOPHOT} and {\tt ALLSTAR} on the galaxy-subtracted master frames and generated ``master star lists'' for further analysis. 
We achieved a photometric precision in our T80S time-series photometry of 0.01~mag or better for objects with $g$, $r$ and $i$ brighter than 16, 15 or 14~mag, respectively. We transformed the T80S measurements into the Pan-STARRS1 photometric system \citep{Tonry2012} and we also calibrated the EABA 1.5m observations in a similar manner.

\subsection{Difference Imaging  Analysis} \label{2kk}

To determine the presence of transients on our surveyed fields, we conducted a difference image analysis (DIA) on them. DIA is a set of methods to perform a photometric comparison among two images of the same portion of the sky.
The idea is to use a reference image that represents the `static sky' and subtract it from the images of interest, revealing a flux excess as result of subtraction. This could be related to new or variable sources. DIA has been applied successfully in many transient searches (The Palomar Transient Factory PTF \citep{PTF_realbogus}, the Nearby Supernova Factory SNFactory \citep{SNFactory}, Pan-STARRS \citep{Chambers2016}, and others).
There are several approaches in the literature \citep{Alard1998}, \citep{Bramich2008}, 
and more recently \cite{Zackay2016}. 
What all these methods have in common is the modeling of a convolution kernel that  minimizes spurious residuals due to differences in instrumentation and atmospheric conditions of each observation, and at the same time unveils the flux variation of true astrophysical sources.
Nonetheless, alignment deformation defects and subtle space PSF variability leave behind a large amount of spurious sources that arise only from the DIA process. 
Dealing with such spurious (or bogus) sources is typically done through human vetting by visual inspection, 
and after this through trained machine classifiers. 
We discuss this topic on section \ref{method}.

For our analysis we implement a Bramich method that fits convolution kernel pixel independently to minimize the L2 norm of the subtraction image.
We obtained  19 images of 17 targets corresponding to observations of GCN \#268556 (GW170104). Each image is a median combination of several individual frames. The images were taken on several nights between the 13th and 21st of January 2017.
The original targets were revisited to obtain references during March 2017 and eventually, due to frosting condensation problems on the CCD,  again on November 2017. The references consist of 17 images, one for each target, and each one a median combination of several individual frames.

\subsection{Real/Bogus classification and\\detection of potential transients} \label{method}

As mentioned before, object detection programs usually detect many DIA artifacts as potential sources. Real transients are commonly outnumbered by these artifacts by a 1:100 ratio or even more. 
Given the large number of these bogus sources, it becomes necessary to train automatic Machine Learning (ML) agents to quickly recognize among the potential candidates the real ones and set them apart from  bogus sources.
ML classifiers are trained with a large number of examples of each class.
Our subtraction images provide us with a large and assorted example of subtraction artifacts (bogus) but we don't have as many examples of real transients, due to the rare nature of these events.

To compensate for the lack of real transients
and to complete a balanced set for real transients, 
we injected synthetic stars on each of the science images, reusing them many times, to achieve equal number of bogus and simulated transients to improve the statistics.
The injected sources were modeled after a PSF estimation of the image done by the \emph{properimage} Python package \citep{Sanchez2018}, based on a Karhunen-Lo\"eve transform of selected stars.
The magnitudes of the injected sources follow the distribution of the ones on the images, as detected by the SEP package \citep{SEPref2016}.

The object detection on the subtraction images was done by the program SExtractor v2.19.5 \citep{Bertin1996}, for objects with a significance over $3.0\sigma$ and with more than 5 connected pixels.
A reduced set of morphological parameters given by this program, was later used as input features to train different ML algorithms.
After recovering the SExtractor parameters of the injected sources we were able to create a labeled (real/bogus) set with 2,414 examples of bogus subtraction artifacts and 2,439 examples of injected transients (4,853 examples in total).
The complete list of features is given in table \ref{tableMLparam}.

We trained a few classical ML algorithms with the training sets obtained. 
The best performance was achieved with a Random Forest algorithm utilizing 10 Decision Trees to avoid over-fitting.
Boosted Decision Trees performed comparably to Random Forest.
The last tested algorithm was Support Vector Machines (SVM), with very low scores at the 60\% level or below.
The final scores are shown on table \ref{tableMLscores}.

The final transient identification was done using the previously trained Random Forest, which swept over the subtraction images, yielding 39 transient candidates (and 2,375 classified bogus).
After visual inspection these were further rejected since they were either misclassified bogus or were not persistent across individual frames.

\begin{figure}
\includegraphics[width=0.5\textwidth]{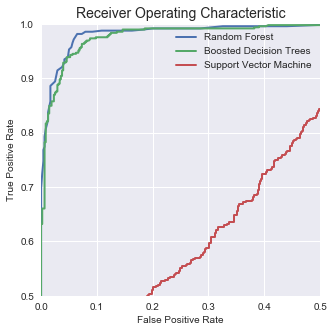}
\caption{The Receiver Operating Characteristic curve for our Random Forest, Boosted Decision Trees and Support Vector Machine classifiers.\label{fig:roccurve}}
\end{figure}

\begin{table*}
\centering
\caption{SExtractor Parameters for Machine Learning\label{tableMLparam} }
\begin{tabular}{lll}
\hline
ID & Parameter \\
\hline
1, 2 & FLUX\_APER, FLUXERR\_APER  \\
3, 4 & MAG\_APER, MAGERR\_APER  \\
5 & FLUX\_MAX \\
6, 7 & ISOAREA\_IMAGE, ISOAREAF\_IMAGE \\
8-13 & X2\_IMAGE, Y2\_IMAGE, XY\_IMAGE, ERRX2\_IMAGE, ERRY2\_IMAGE, ERRXY\_IMAGE \\
14-19 & CXX\_IMAGE, CYY\_IMAGE, CXY\_IMAGE, ERRCXX\_IMAGE, ERRCYY\_IMAGE, ERRCXY\_IMAGE \\
20-25 & A\_IMAGE, B\_IMAGE, THETA\_IMAGE, ERRA\_IMAGE, ERRB\_IMAGE, ERRTHETA\_IMAGE \\
26-33 & ISO0, ISO1, ISO2, ISO3, ISO4, ISO5, ISO6, ISO7 \\
34 & FLAGS \\
34 & FWHM\_IMAGE \\
36 & ELONGATION \\
37 & ELLIPTICITY \\
38 & POLAR\_IMAGE \\
39, 40 & VIGNET, VIGNET\_SHIFT \\
41 & FLUX\_GROWTHSTEP \\
42, 43 & MAG\_GROWTH, MAG\_GROWTHSTEP \\
44 & FLUX\_RADIUS 
\end{tabular}
\end{table*}

\begin{table*}
\centering
\caption{Machine Learning Scores \label{tableMLscores}}
\begin{tabular}{lccc}
\hline
 & Random Forest & Boosted Decision Trees & Support Vector Machine \\
 \hline
Accuracy & 0.89 & 0.89 & 0.65 \\
Precision & 0.92 & 0.91 & 0.79  \\
Recall & 0.86 & 0.87 & 0.48 \\
F1-measure & 0.89 & 0.89 & 0.59 \\
\hline
\end{tabular}
\end{table*}

\subsection{Testing Convolutional Neural Network}

Although our results show the Random Forest (RF) method as the optimal method to find optical transients in taken images we  still investigated possible more efficient methods. 
In this paragraph we describe our experimental real/bogus classification method for image subtraction that is based on Convolutional Neural Networks (CNN). In general, CNN-based methods are used in image processing and construction of the network is inspired by real organism vision systems. Classification of real/bogus events is also an image processing problem. The method works in a similar manner as the RF algorithm on subtracted images, searching for potential OTs that might came up during the subtraction process. But CNN handle things somehow differently: it uses as input a cropped part of the subtraction  image which contains the potential source selected by SExtractor, and not a set of parameters associated to the potential source selected by SExtractor. 

The CNN method is as follows. First SExtractor finds all potential sources on the subtracted image above a 3$\sigma$ detection threshold. Then, a 28 by 28 pixel area around a source is cropped from the image and fed to a neural network to label it as a real transient or bogus (artifact of the subtraction process). After getting the likelihood of the potential OT belonging to one those classes, we identify its class to the class that scored higher in the classification. The construction of the network is described in table \ref{tab:cnn}. We trained and tested a CNN network  the same manner as we did for the RF, except the input to the CNN are the raw pixels of the cropped images, as opposed to SExtractor parameters in the RF case. Our network achieved ~99.5\% accuracy on testing and training sets, matching all of the cases\footnote{If the dropout layer between two dense layers has the parameter set to 0.6 it is possible to achieve 100\% accuracy.}. After running on all data connected to GW170104 we did not find any candidate for OT, which is consistent with RF method and human inspection of the best candidates.

To build our neural network, we used the Keras package with a Theano backend. We chose a Sequential type model with \texttt{categorical\_crossentropy} loss metric. We used the `Adam' optimizer and accuracy metrics. Our construction of the network is a bit different from other networks like that of \cite{CNN2016,CNN2017}. In our approach we stacked two Conv2D layers together and used high dropout values at the dense part of the network. It is important to note that our experimental solution offer better accuracy than \cite{CNN2016} or \cite{CNN2017}. Our method based on CNNs will be subject to future development and will be expanded in a separate report.

\begin{table}
\caption{Construction of experimental Convolutional Neural Network used to real/bogus classification}
\label{tab:cnn}
\begin{tabular}{ll}
Layer Name & Parameters \\
\hline
Convolution2D &(32, 3, 3, activation='relu', input\_shape=(1,21,21)) \\
Convolution2D &(32, 3, 3, activation='relu') \\
MaxPooling2D &(pool\_size=(2,2)) \\
Dropout &(0.25) \\
Flatten &() \\
Dense &(128, activation='relu') \\
Dropout &(0.57) \\ 
Dense &(2, activation='softmax')
\end{tabular}
\end{table}

\section{Results}\label{results}
\subsection{GW170817 - SSS17a}

\begin{figure*}
\includegraphics[width=0.4\textwidth,angle=0]{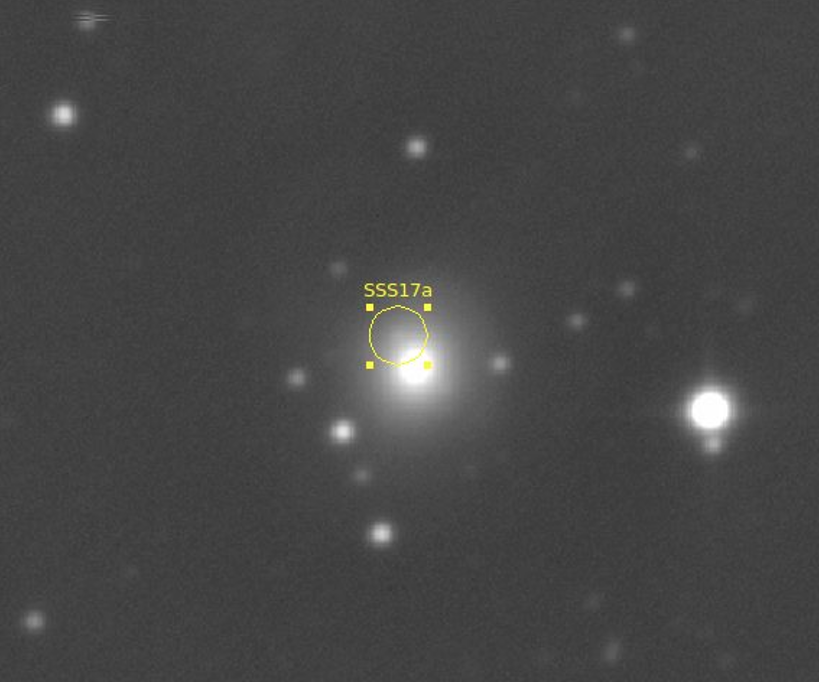}
\includegraphics[width=0.39\textwidth,angle=0]{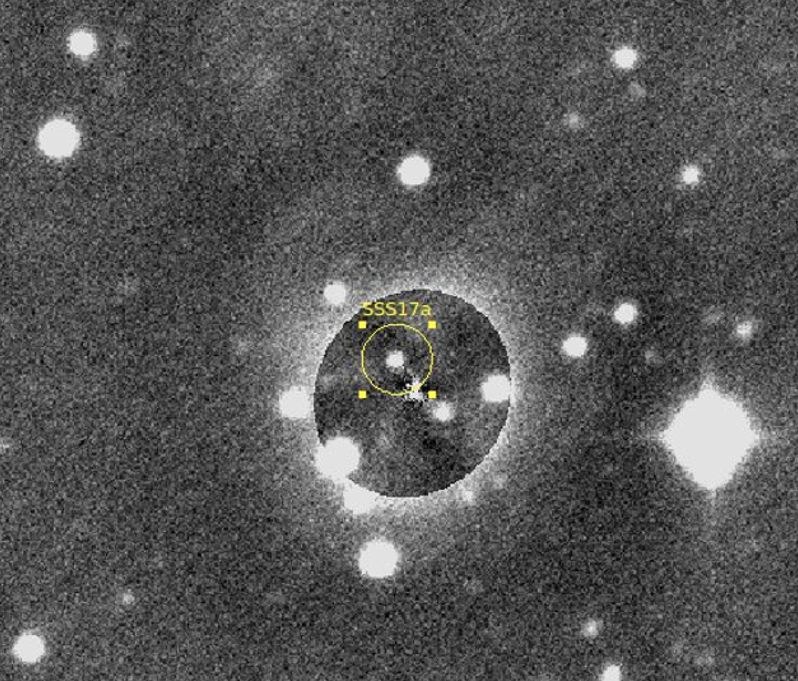}
\caption{Left Image: clear filter image of NGC 4993 centered on the transient, combined from images taken by EABA 1.54m telescope. Intensity scaled logarithmically to better show the light distribution of the host galaxy. Right Image: same as previous image after host galaxy was subtracted. \label{fig:eaba}}
\end{figure*}

The TOROS Collaboration also participated in the search for an optical counterpart of GW170817. The main results are described in \cite{Diaz2017}. This section  summarizes our findings. Our follow-up started just few hours after the trigger. On the first two nights 17$^{th}$-18$^{th}$ August 2018 we have imaged 26 galaxies potentially linked to GW170817 that have been selected by our target selection algorithm. The initial search was done using two facilities, the T80-South telescope  and a Meade LX200 16-inch telescope equipped with a SBIG STF 8300 camera, located in Tolar Grande, Argentina \citep{GCN21619,GCN21620}. When the candidate counterpart was identified near NGC$\,$4993 our follow-up was focused on this source. On 18$^{th}$ August the transient was observed by T-80 South that took 16, 15, and 15 one-minute exposures through SDSS $g$, $r$ and $i$ filters, respectively, over a course of 80 minutes. The light curve in all bands exhibited a very significant decline across all bands during the $\sim 80$~minutes of observations. Figure \ref{fig:eaba} shows a transient as observed by T-80 South telescope with and without galaxy subtraction. On 19$^{th}$ of August observations were continued using the 1.54m telescope located at the Estaci\'on Astrof\'{\i}sica de Bosque Alegre (EABA). The observations taken with EABA were done without using any astronomical filter. However the extrapolation of values of brightness in $r$ filter that we obtained, was consistent with the decline trend. Our observation of the light curve of the transient, was inconsistent with a ``red kilonova'' model. Our `r-band' light-curve is in fair agreement with a ``blue kilonova'' model \citep{Metzger2017} with ``wind'' free of lanthanides, but predicted luminosities are lower. More information about our observations could be found in \citep{Diaz2017} or observations done by other teams \citep{Abbott2017}.

\subsection{GW170104}

No transient was found. The Machine Learning algorithm found about 50 candidates, but after human inspections there were all rejected. We assume that there was no bona fide candidate for electromagnetic transient present near to any galaxy that we have observed. It has to be pointed out that this is consistent with the current state of knowledge, were binary black hole mergers are not likely to produce an EM counterpart. 

\subsection{Remarks about strategy and data analysis\label{sec:strategy}}

The event GW170817 makes us reconsider the targeting strategy for future events. Our first approach for galaxy targeting  use very restrictive cuts  on absolute and apparent blue magnitude of the galaxy ($< -17.5, < 19.0  $ respectively), distance ( $< 80 MPC$) and relative position to the sun that the object would have a zenith distance of 45 deg or lower. A list after those cuts was sorted based on its likelihood and the top 20 galaxies were considered. GW170817 did not make it to that list on the first night since it was never 45 deg above the horizon for any of our observatories. If there will not be any cuts on the zenith distance GW170817 would have ranked 25$^{th}$. 

After GW170817 we revised our strategy for choosing targets. First of all, cuts rejecting galaxies must not be too conservative, they should be as low as possible giving a chance for operators and AI systems at the observatories, to reject galaxies that fall below the horizon. Secondly, from event GW170817 we can assume that with very high probability, optical transients connected with a kilonova will occur next to massive galaxies. In the area of the LIGO final skymap where probability was higher than $10^{-6}$ per skymap's pixel, NGC 4993 was the sixth most massive galaxy. We checked few different strategies for choosing target galaxies in the upcoming O3 science run. The best strategy consists of following a ranking statistics along with a product of the likelihood and the absolute magnitude of the galaxy. In this ranking statistics, NGC 4993 comes in 6$^{th}$ place. Galaxies are only rejected based on the possibility of observation from the current location, the cut-off  position above the horizon is based on requests from the given observatory. This strategy will be subject to future testing in the O3 science run.

\section{The TOROS observatory at Cord\'{o}n Mac\'{o}n}

\begin{figure*}
\includegraphics[width=0.8\textwidth,angle=0]{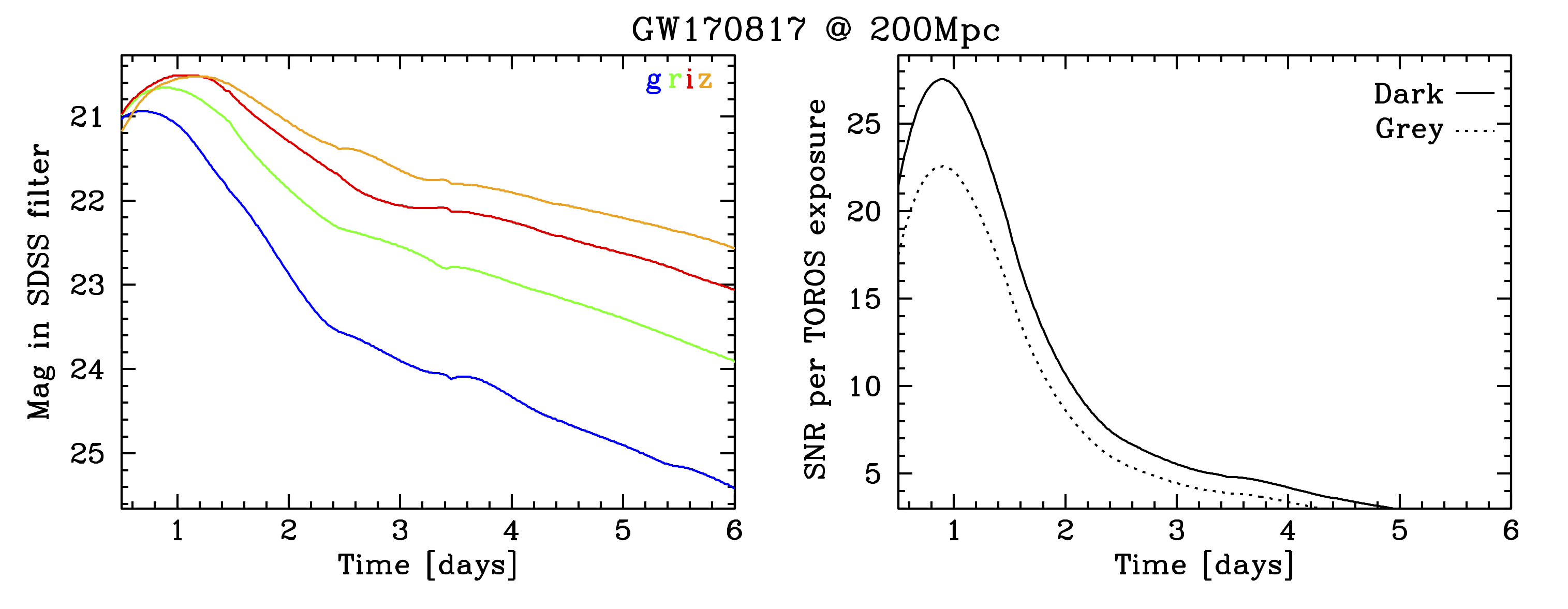}
\caption{Left Image: plot shows the light curves of GW170817 \citep{Cowperthwaite2017},
shifted to a distance of 200 Mpc. Right Image: plot shows the expected SNR of TOROS photometry
of such an event. "Dark" and "grey" refer to typical sky brightness values associated with
the new and quarter Moon, respectively. The system should be
sensitive (SNR> 3) to kilonova events for at least 4 days.\label{fig:torosNew}}
\end{figure*}

To participate in the upcoming  O3 LIGO VIRGO observational campaign we are currently installing a system with a primary mirror diameter of 0.6m which will in its final design stage have a large field-of-view (10 sq.deg.) camera with a very broad bandpass (0.4-0.9 $\mu$m, equivalent
to a combination of the Sloan \emph{griz} filters). The system will be installed atop Cord\'{o}n Mac\'{o}n located in the Atacama portion of Northwestern Argentina. This dedicated instrument to follow-up GW candidate events in the Southern Hemisphere will fill a niche and offer extended coverage of the southern skies. The telescope (Planewave CDK24 with LT500 mount) and its dome (Ashdome Lanphier type) are currently undergoing installation at the site. We plan to  equip the telescope with a 10K$\times$10K backside-illuminated STA1600LN CCD and prime-focus  corrector to maximize the field of view of the telescope.  It is worth  noting that even in below average conditions, TOROS remains sensitive (SNR> 3) to kilonova events for at least 4 days (see Fig. \ref{fig:torosNew}). 

\section{Summary}\label{conclusions}

The TOROS collaboration conducted a prompt search for EM counter parts to gravitational wave events reported by LIGO-Virgo detectors during their second science run (O2) using the 1.5-m telescope of Estaci\'on Astrof\'{\i}sica Bosque Alegre (EABA) in C\'ordoba, Argentina and  T80-South (T80S) at the Cerro Tololo Inter-American Observatory, Chile.

We followed-up three LIGO alerts GW170104 (BBH), GW170120 (BBH) and GW170817 (NS-BH). For two of them GW170104 and GW170817 we performed detailed analysis searching for optical counter part. For GW170817 the optical counter part was identified and light curves were derived. The data gathered by the TOROS team were an important part of the worldwide observations of the first joint observed event in both gravitational waves and electromagnetic spectrum. It was also the first kilonova ever observed. For GW170104 we did not found any optical counterparts. However this was expected due to the detection of a binary black hole merger as origin of the event. For GW170120, the data gathered did not have enough quality to perform useful scientific analysis.

The discovery of the optical counterpart to GW170817 in close proximity to NGC4993 proves that the strategy used by to identify EM counterparts to gravitational wave transient was appropriate. The strategy, a galaxy-based approach, focuses the search on the most massive galaxies in the probability area of origin of the gravitational wave. Using this approach NGC4993 ranked in 30 galaxies to image.

The TOROS network is expanding and for O3 LIGO-Virgo Science Run we expect to have two fully dedicated sites for fast transient search. One in Salta, Argentina and one in Brownsville, Texas. Both observatories will be fully robotic. In addition, the collaboration has strengthened their capabilities by  observation time allocated at the Gran Telescopio de Canarias (20-MULTIPLE-3-19AMEX, PI Omar L\'{o}pez-Cruz),  Gemini Sur (GS-2019A-Q-122, PI Diego G. Lambas) and the FALCON Telescope Network(Chun et al. 2018) (FTN-MMO-20180010, PI Nilo Castellon).

{\it The TOROS collaboration acknowledges support from Ministerio de Ciencia, Tecnolog\'{\i}a e Innovaci\'on Productiva (MinCyT) and Consejo Nacional de Investigaciones Cient\'{\i}ficas y Tecnol\'ogicas (CONICET) from Argentina, grants from the National Science Foundation of the United States of America, NSF PHYS 1156600 and NSF HRD 1242090, and the government of Salta province in Argentina. The T80-South robotic telescope (Mendes de Oliveira et al., submitted)
was founded as a partnership between the S\~{a}o Paulo Research Foundation
(FAPESP, grant number 2009/54202-8), the Observat\'{o}rio Nacional (ON), the Federal University of
Sergipe (UFS) and the Federal University of Santa Catarina (UFSC), 
with important financial and practical contributions from other
collaborating institutes in Brazil, Chile (Universidad de La Serena)
and Spain (CEFCA). JLNC is grateful for financial support received from the GRANT PROGRAMS FA9550-15-1-0167 and FA9550-18-1-0018 of the Southern Office of Aerospace Research and development (SOARD), a branch of the Air Force Office of the Scientific Research International Office of the United States (AFOSR/IO). NCNR is grateful for financial support from MNiSW grant DIR/WK/2018/12.}

\ \par

{\it Facilities:} EABA, T80-S, TORITOS.

\bibliographystyle{apj}


\bsp	
\label{lastpage}
\end{document}